\documentclass[pre,aps,twocolumn,showpacs,superscriptaddress,floatfix]{revtex4}
\input epsf

\usepackage{amssymb}
\usepackage{graphicx}
\usepackage{amsmath}
\usepackage{latexsym}
\usepackage{verbatim}

\begin{document}

\title{Critical behavior of the Ising model in annealed scale-free networks}

\author{Sang Hoon Lee}
\affiliation{Department of Physics,
Korea Advanced Institute of Science and Technology, Daejeon 305-701, Korea}

\author{Meesoon Ha}
\affiliation{Department of Physics, Korea
Advanced Institute of Science and Technology, Daejeon 305-701, Korea}
\affiliation{School of Physics, Korea Institute for Advanced Study,
Seoul 130-722, Korea}

\author{Hawoong Jeong}
\affiliation{Department of Physics,
Korea Advanced Institute of Science and Technology, Daejeon 305-701,
Korea} \affiliation{Institute for the BioCentury, Korea Advanced
Institute of Science and Technology, Daejeon 305-701, Korea}

\author{Jae Dong Noh}
\affiliation{Department of Physics, University of Seoul, Seoul 130-743, Korea}

\author{Hyunggyu Park}
\affiliation{School of Physics, Korea Institute for Advanced Study,
Seoul 130-722, Korea}

\date{\today}

\begin{abstract}

We study the critical behavior of the Ising model in annealed
scale-free~(SF) networks of finite system size with forced upper cutoff in
degree. By mapping the model onto the weighted fully connected Ising model,
we derive analytic results for
the finite-size scaling~(FSS) near the phase transition,
characterized by the cutoff-dependent two-parameter
scaling with four distinct scaling regimes,
in highly heterogeneous networks. These results are
essentially the same as those found for the nonequilibrium
contact process in annealed SF
networks, except for an additional complication due to the trivial
critical point shift
in finite systems. The discrepancy of the FSS theories between annealed and
quenched SF networks still remains in the equilibrium Ising model,
like some other
nonequilibrium models. All of our analytic results are confirmed reasonably
well by numerical simulations.
\end{abstract}


\maketitle

\section{introduction}
Many aspects of our real world have been understood in the context of
complex networks~\cite{NetworkReview,DorogovtsevBook} and
simple physical models of critical phenomena on networks.
Contrary to regular lattices in the Euclidean space,
complex networks are characterized by a highly heterogeneous structure
as manifested in broad degree distributions.
Recent studies on equilibrium or
nonequilibrium systems have revealed that the heterogeneity is one of
essential ingredients determining the universal
feature of phase transitions and critical phenomena~\cite{Dorogovtsev2007}.

The concept of the phase transition is well defined only in the thermodynamic
limit where the system size is taken to infinity. So it is important
to understand how finite-size effects come into play near the transition.
Such a task for physical models on regular lattices has been successfully
accomplished by the standard finite-size scaling~(FSS) theory~\cite{Privman1990},
based on the ansatz that a single characteristic length
scale~(correlation length) $\xi$
competes with the system's linear size $L$. Then, any physical observable
depends only on a dimensionless variable $\ell=L/\xi$ in the scaling limit.
Near a second-order continuous transition, the correlation length diverges as
$\xi\sim |\epsilon|^{-\nu}$ with the reduced coupling constant $\epsilon$ and
the finite-size effects become prominent.

The FSS theory for complex networks can be formulated in a similar way:
Since the Euclidean distance is undefined in complex networks,
one may take the volume scaling variable as ${\ell}_v = N/\xi_v$
with the system size $N$~(the total number of nodes)
and the correlated volume $\xi_v$.
The correlated volume diverges
$\xi_v\sim |\epsilon|^{-\bar{\nu}}$ near the
transition~($\bar\nu = \nu d$ in $d$ dimensional lattices).
For example, the magnetization  of the Ising model scales as
\begin{equation}
\label{fss}
m(\epsilon,N)=N^{-\beta/\bar\nu} \psi (\epsilon N^{1/\bar\nu}),
\end{equation}
where the scaling function $\psi(x)\sim O(1)$ for small $x$
and $x^\beta$ for large $x$
with the order parameter exponent $\beta$.

The FSS theory with a single characteristic size has been tested
numerically in many systems~(see Ref.~\cite{Dorogovtsev2007}
and references therein). In particular,
the exact values for the FSS exponent $\bar{\nu}$ are
conjectured~\cite{HHong2007}
by estimating the correlated volume~(droplet) size
for the nonequilibrium contact process~(CP) and the equilibrium Ising model
in random uncorrelated networks with static links,
which are denoted as {\em quenched} networks.

However, considering a highly heterogeneous scale-free~(SF) network, one
should take into account not only a broad degree distribution of
$P(k) \sim k^{-\lambda}$ but also
the upper cutoff $k_c$ in degree, which scales as $k_c\sim N^{1/\omega}$.
Without any constraint, $k_c$ is bounded naturally with
$\omega_{\rm nat}=\lambda-1$.  In general, one may impose
a {\em forced} cutoff with $\omega>\omega_{\rm nat}$.
In the thermodynamic limit, both $N$ and $k_c$ diverge simultaneously
and $\omega$ sets a route to the limit. Therefore, one can suspect that
the FSS theory may depend on the routes or equivalently on the value of
$\omega$, especially for networks with a broader distribution
for small $\lambda$.

For the quenched SF networks, it has been suggested that the FSS does not vary
with $\omega$ for a weak forced cutoff~($\omega<\lambda$), which was confirmed
numerically in various types of SF networks~\cite{HHong2007,Ha2007}. However,
in the {\em annealed} networks where links are not fixed but fluctuate randomly in time,
it was rigorously shown that the CP model exhibits an anomalous FSS for any forced cutoff
 with $2<\lambda<3$ where a heterogeneity($\lambda$)-dependent
critical scaling appears~\cite{CP-S2007,Castellano2007,Boguna2009,JDNoh2009}.
Moreover, the anomalous FSS is characterized by a
cutoff($\omega$)-dependent and two-parameter scaling with four distinct scaling regimes~\cite{JDNoh2009},
in contrast to the cutoff-independent and single-parameter scaling with three scaling
regimes in the standard FSS theory.

The anomalous FSS of the CP in the annealed SF networks gives rise to a
natural question: What is the main ingredient causing the anomaly?
Some possible guesses may be a nonequilibrium feature
of the CP,  absorbing nature~(vanishing activity) at criticality,
or heterogeneity of networks~\cite{Castellano2007,Boguna2009}.
In this paper, we answer to this question by studying
the Ising model, a prototype equilibrium phase transition model,
in annealed SF networks.
We find the same type of the anomalous FSS scaling~(cutoff-dependent
and two-parameter scaling with four distinct scaling regimes)
for any forced cutoff with $3<\lambda<5$ where the $\lambda$-dependent critical scaling
appears in the thermodynamic limit for the Ising version. In addition, the
trivial shift of the critical point in finite systems adds one more complication on
the critical FSS, though it does not cause any fundamental change.
In summary, our results may draw a general conclusion
that the anomalous FSS scaling should appear in any critical system in the annealed SF
networks for any forced cutoff~($\omega>\omega_{\rm nat}$)
with the degree exponent $\lambda$ such that a $\lambda$-dependent new singularity arises
in the physical quantities as $N\rightarrow\infty$.

This paper is organized as follows. We define the Ising model on an annealed
network in Sec.~\ref{sec2} and show that it is equivalent
to the Ising model on the weighted fully connected network. In Sec.~\ref{sec3},
the FSS theory is developed in various networks including SF networks,
which is numerically tested in Sec.~\ref{sec4}. In Sec.~\ref{sec5},
some effects of the sampling disorder are discussed. We conclude this paper
with summary and discussion in Sec.~\ref{sec6}.

\section{Ising model on annealed networks}
\label{sec2}

An annealed network $\mathcal{G}_N$ is defined as an ensemble of all networks
consisting of $N$ nodes which are assigned to a given degree
sequence $\{k_1,\ldots,k_N\}$. An instance $g\in \mathcal{G}_N$ is
constructed by assigning $k_i$ stubs to each node $i~(1,\ldots,N)$ and then
completing edges by pairing the stubs randomly as in the uncorrelated configuration
model~\cite{Newman2001,CBP-S2005}.

A network configuration $g$ is
conveniently represented by an adjacency matrix $A(g)$ whose element
$A_{ij}$ takes either 1 or 0 if there is an edge between nodes $i$ and
$j$ or not, respectively.
In the ensemble $\mathcal{G}_N$, the connecting probability
$p_{ij}$ to find an edge between two nodes $i$ and $j$ is given by~\cite{Dorogovtsev2007,PN2004}
\begin{equation}\label{pij}
p_{ij} = \frac{ k_i k_j}{N z_1} + \mathcal{O}\left(\frac{1}{N^2}\right) ,
\end{equation}
with the mean degree $ z_1 \equiv \sum_ik_i / N$. This expansion is valid
when $\frac{ k_i k_j}{N z_1}\ll 1$ for all $i$ and $j$.

The ferromagnetic Ising model on the annealed network $\mathcal{G}_N$ is
defined by the Hamiltonian
\begin{equation}
H[\{s\},g] = - {J} \sum_{i<j} A_{ij}(g) s_i s_j -\sum_i h_i s_i ,
\end{equation}
where ${J}>0$ is a ferromagnetic coupling constant,
$s_i \in \{-1,1\}$  is an Ising spin variable at node $i$, and
$h_i$ is a local field at node $i$.
In comparison to
the model on a quenched network, a network configuration $g$ is also
fluctuating within $\mathcal{G}_N$ as well as the Ising spins.
Thermodynamic properties of the model is obtained from the partition
function
\begin{equation}
Z = \sum_{g\in\mathcal{G}_N} \sum_{\{s_i\}}
\exp\left[{K} \sum_{i<j} A_{ij}(g) s_i s_j+\sum_i \tilde{h}_i s_i \right] ,
\end{equation}
where ${K} =\beta {J}$ and $\tilde{h}_i=\beta h_i$ with
the inverse temperature $\beta=1/k_BT$.

In terms of the connection probability in Eq.~(\ref{pij}),
one can easily perform the partial summation over $g$ to obtain that
\begin{equation}
Z = \sum_{\{s_i\}} \prod_{i<j} \left[ (1-p_{ij}) + p_{ij} e^{{K} s_i
s_j} \right]\prod_i e^{\tilde{h}_i s_i} .
\end{equation}
Utilizing the identity $e^{{K} s} = \cosh {K} +
s \sinh {K}$ for $s=\pm 1$, we find that
\begin{equation}\label{Z}
Z = Z_0 \sum_{\{s_i\}} \exp \left( \sum_{i<j} Q_{ij} s_i s_j +\sum_i \tilde{h}_i s_i \right) ,
\end{equation}
where $Z_0$ is an overall constant factor (not depending on $\{ s_i \}$),
$$
Z_0 = \prod_{i < j} \left( \frac{1 - p_{ij} + p_{ij} \cosh K}{\cosh Q_{ij}} \right) ,
$$
and
\begin{equation}
\tanh Q_{ij} =  \frac{ p_{ij} \sinh {K}}{ 1 - p_{ij} + p_{ij} \cosh {K}} \ .
\end{equation}
As $Q_{ij}$ is nonzero for any pair of $(i,j)$, the
expression in Eq.~(\ref{Z}) corresponds to the partition function of
the Ising model on the fully connected network with the heterogeneous
coupling constants $Q_{ij}$.

As $p_{ij} = k_ik_j / (N z_1) \ll 1$ for large $N$~\cite{exp1}, one can approximate
$Q_{ij} \simeq \tilde{K} k_i k_j / ( N z_1)$ with
$\tilde{K} = \sinh {K}$. Hence, in this paper, we focus on studying
the Ising model on the fully connected network with the Hamiltonian
$H_f$ as
\begin{equation}\label{H_s}
\beta H_f = -\tilde{K} \sum_{i<j} \frac{k_i k_j}{N z_1} s_i s_j -\sum_i \tilde{h}_i s_i .
\end{equation}
This Hamiltonian was studied as a MF or annealed approximation for the Ising model on
quenched networks in the thermodynamic limit~\cite{Dorogovtsev2007,Bianconi2002,Igloi2002}.

For convenience, we rewrite $H_f$ in a completed square form as
\begin{equation}
\beta H_f = -\frac{\tilde{K}}{2 N z_1} \left[ \left(
\sum_i k_i s_i \right)^2 - \sum_i k_i^2 \right] -\sum_i \tilde{h}_i s_i\
\label{Hs}
\end{equation}
and define the magnetic order parameter as
\begin{equation}
\tilde{M} \equiv \sum_i k_i s_i ,
\end{equation}
with the order parameter density $\tilde{m} \equiv \tilde{M}/(N z_1 )$,
which is first suggested in \cite{Dorogovtsev2003} and recently
for both equilibrium and nonequilibrium models in \cite{Caccioli2009}.

Now we derive the free energy as a function of $\tilde{M}$, which allows us
to calculate thermodynamic properties
not only in the thermodynamic limit by minimizing it with respect to
$\tilde{M}$ but also for finite size $N$, at least up to the leading order.
After dropping the additive constant term
in Eq.~(\ref{Hs}),
the partition function, up to a constant, can be written as
\begin{eqnarray}\label{partition}
Z &=& \sum_{\{s_i\}} \exp\left[ \frac{\tilde{K}}{2Nz_1}
\left( \sum_i k_i s_i\right)^2  +\sum_i \tilde{h}_i s_i \right] \nonumber \\
  &=& \int d\tilde{M} \sum_{\{s_i\}} e^{\tilde{K}\tilde{M}^2/(2Nz_1)+\sum_i \tilde{h}_i s_i}
\delta\left(\tilde{M}-\sum_i k_i s_i\right) \nonumber \\
 &=& \int d\tilde{M} \int_{-i\infty}^{i\infty} \frac{du}{2\pi i} \exp[- \tilde{F}(\tilde{M},u) ] \ ,\label{z_m}
\end{eqnarray}
where
$$
\tilde{F}(\tilde{M},u)
\equiv -\frac{\tilde{K}}{2Nz_1} \tilde{M}^2 + u\tilde{M} -
N\overline{ \ln [2\cosh (uk_i+\tilde{h}_i)]} ,
$$
where $\overline{()_i} \equiv \frac{1}{N}\sum_{i}()_i$ denotes the average
over nodes.
In obtaining Eq.~(\ref{z_m}), we used the integral representation of the
delta function $\delta(\tilde{M}) = \int\frac{dv}{2\pi } e^{iv\tilde{M}}$ and the analytic
continuation $v=iu$.

The integration over $u$ can be evaluated using the steepest
descent method, which yields that the free-energy function $F(\tilde{M})$ defined by
$Z\equiv\int d\tilde{M} \exp(-F(\tilde{M}))$
is given as
\begin{equation}\label{F_M}
F(\tilde{M}) \simeq \tilde{F}(\tilde{M},u_0) +\frac{1}{2} \ln
\left[2\pi |\tilde{F}^{\prime\prime}|\right] +\cdots ,
\end{equation}
where $\tilde{F}^{\prime\prime}=-N\overline{ k_i^2 {\rm sech}^2 (u_0k_i+\tilde{h}_i)}$ is the partial second derivative of $\tilde{F}(\tilde{M},u)$ with respect to $u$ at $u_0$. The condition that the first derivative
$\tilde{F}^\prime|_{u=u_0}=0$ determines $u_0=u_0(\tilde{M},\{\tilde{h}_i\})$ by
\begin{equation}
\tilde{M} =  N \overline{k_i \tanh (u_0 k_i +\tilde{h}_i)}
\end{equation}
or equivalently, $u_0=u_0(\tilde{m},\{\tilde{h}_i\})$ by
\begin{equation}\label{s_c}
\tilde{m} = \frac{1}{z_1 } \overline{ k_i \tanh (u_0 k_i +\tilde{h}_i)} \ .
\end{equation}
We remark that the second and high-order terms on the right hand side of
Eq.~(\ref{F_M}) can be neglected because they increase with system size $N$ only logarithmically,
in contrast to the first bulk term. Moreover, the finite-size corrections near the
transition are stronger than the contributions from these terms.

It is convenient to use the free energy density function $f(\tilde{m})\equiv F(\tilde{M})/N$
which is
\begin{equation}
f(\tilde{m})\simeq -\frac{z_1\tilde{K}}{2} \tilde{m}^2 + z_1 u_0 \tilde{m}-
\overline{ \ln [2\cosh (u_0 k_i+\tilde{h}_i)]} .
\end{equation}
Then, the ensemble-averaged value $\langle\tilde{m}\rangle$ can be calculated,
in the thermodynamic limit, as
\begin{equation}\label{finite}
\langle\tilde{m}\rangle = \frac{1}{\tilde{Z}}\int d\tilde{m}\ \tilde{m} e^{-Nf(\tilde{m})}\approx
\tilde{m}_0 ,
\end{equation}
where $\tilde{Z}=Z/(Nz_1)=\int d\tilde{m}\ e^{-N f(\tilde{m})}$ and
$\tilde{m}_0$ is the minimum point of $f(\tilde{m})$. Here,
the higher-order finite-size corrections are again at most logarithmic.

The spin magnetization $m_i$ at node $i$ can be obtained by differentiating
the partition function in Eq.~(\ref{partition}) by the local field $\tilde{h}_i$,  which result in
\begin{eqnarray}\label{relation}
m_i=\langle s_i\rangle&=&\left\langle \tanh (uk_i+\tilde{h}_i) \right\rangle \nonumber\\
&\simeq& \tanh \left[\tilde{u}_0k_i +\tilde{h}_i\right],
\end{eqnarray}
with $\tilde{u}_0\equiv u_0(\tilde{m}_0,\{\tilde{h}_i\})$.

\section{FSS theory in annealed networks}
\label{sec3}

We are now ready to investigate the bulk critical scaling and also the
FSS of the Ising model on annealed networks.
First, we consider the simplest case of exponential
degree distributions such as the Poisson distribution of the random network.
Then, we proceed to discuss for the SF degree distributions with
$P(k)\sim k^{-\lambda}$ with an upper cutoff $k_c\sim N^{1/\omega}$.

\subsection{Exponential networks}

Consider exponentially bounded degree distributions such that
the degree moments $z_n\equiv \overline {k_i^n}$ are bounded for
all $n$. The Poisson distribution for the random network and
the Kronecker $\delta$-function distribution [$P(k) = \delta_{k,z}$]
for the random $z$-regular network fall into this
category.

Taking the uniform magnetic field $\tilde{h}_i = \tilde{h}$ and expanding
Eq.~(\ref{s_c}) for small $u_0$ and $\tilde{h}$, we get
\begin{equation}
\label{m_u0_ll}
\tilde{m} = \tilde{h}+\frac{z_2}{z_1} u_0 - \frac{z_4}{3z_1} u_0^3 +
\mathcal{O}\left(u_0^5,\tilde{h}^2, \tilde{h}u_0^2\right)  .
\end{equation}
Then, the free energy density is given by
\begin{equation}\label{freg}
f(\tilde{m}) = -\ln 2 -a \tilde{h}\tilde{m}-
\frac{a}{2}\epsilon \tilde{m}^2
+ \frac{b}{12} \tilde{m}^4 + \cdots \ ,
\end{equation}
where
\begin{equation}\label{def_a_b}
a = {z_1^2}/{z_2} , \quad b = {z_1^4 z_4}/{z_2^4} ,
\end{equation}
and the reduced inverse temperature $\epsilon=(\tilde{K}-\tilde{K_c})/\tilde{K}_c$ with
the critical point
\begin{equation}\label{def_Kc}
\tilde{K}_c=z_1/z_2 .
\end{equation}
Note that $a=b=1$ and $\tilde{K_c} = 1 /z$ for
the random $z$-regular networks.

At $\tilde{h}=0$, the order parameter scales for $\epsilon>0$ as
\begin{equation}
\label{exp:m_below_tc}
\langle\tilde{m}\rangle \simeq \sqrt{3a/b} \ \epsilon^\beta,
\end{equation}
with the order parameter exponent $\beta=1/2$.
It is straightforward to derive the zero-field susceptibility $\tilde{\chi}
\equiv \partial \langle \tilde{m}\rangle / \partial\tilde{h} |_{\tilde{h}=0}
\simeq (2\epsilon)^{-\gamma}$ for $\epsilon>0$ and $\tilde{\chi} \simeq
(1-2/\pi)(-\epsilon)^{-\gamma}$ for $\epsilon<0$ with $\gamma=1$.
The average magnetization $m\equiv \overline{m_i}$ is related to $\langle\tilde{m}\rangle$
through Eq.~(\ref{relation}), which yields $ m\simeq a\langle\tilde{m}\rangle$.

With the free energy function given in Eq.~(\ref{freg}), one can develop the
FSS theory analytically. The full scaling functions for
$\langle\tilde{m}\rangle$ and $\tilde{\chi}$ are derived in the
Appendix.
We only summarize the results below. The FSS form for the
order parameter is given by
\begin{equation}\label{m_fss}
\langle \tilde{m}(\epsilon,N) \rangle = N^{-\beta/\bar{\nu}}
\tilde{\psi}(\epsilon N^{1/\bar{\nu}} ; a,b) ,
\end{equation}
where $\beta=1/2$, the FSS exponent $\bar{\nu}=2$,
and the scaling function $\tilde{\psi}$ is
given by Eq.~(\ref{m_s_f}). The function arguments $a$ and $b$
will be omitted from now on unless it causes confusion.

\begin{figure}
\includegraphics[width=0.9\columnwidth]{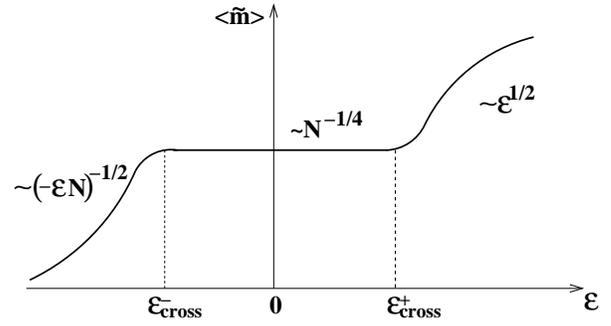}
\caption{Schematic plot of $\langle\tilde{m}\rangle$
versus $\epsilon$ in the exponential networks.}
\label{fig1}
\end{figure}

The critical FSS at $\epsilon=0$ is
\begin{equation}
\label{exp:m_tc}
\langle\tilde{m}\rangle_c \simeq A_e N^{-1/4},
\end{equation}
where $A_e=\tilde{\psi}(0) =
(12/b)^{1/4}\Gamma(\frac{1}{2})/\Gamma(\frac{1}{4})$.
We remark that $\epsilon$ may not be exactly zero at the bulk critical point
$\tilde{K}_c^\infty=\lim_{N\rightarrow\infty} \tilde{K}_c$,
but may have a finite-size correction vanishing
exponentially with $N$. This additional correction does not change the leading
power-law term in the FSS.
For $\epsilon<0$,
\begin{equation}
\langle\tilde{m}\rangle\simeq \sqrt{2/(\pi a)}\ (-\epsilon N)^{-1/2} ,
\end{equation}
since $\tilde{\psi}(x) \simeq \sqrt{-2x/(\pi a)}$ for $x \rightarrow -\infty$.
The scaling form in Eq.~(\ref{exp:m_below_tc})
is reproduced from Eq.~(\ref{m_fss}), using the limiting behavior of
$\tilde{\psi}(x) \simeq \sqrt{3ax /b}$ for $x\rightarrow \infty$.
The crossover between the three scaling regimes occurs at
\begin{equation}
\epsilon_{\rm cross}^- \simeq -[2/(\pi aA_e^2)]N^{-1/2} ,
\label{e_cross_-}
\end{equation}
and
\begin{equation}
\epsilon_{\rm cross}^+ \simeq [bA_e^2/(3a)]N^{-1/2} .
\end{equation}
The scaling behavior of the order parameter $\langle \tilde{m}\rangle$ is
represented schematically in Fig.~\ref{fig1}.

The FSS form for the zero-field susceptibility is given by
$\tilde{\chi} = N^{\gamma/\bar{\nu}} \tilde{\phi}(\epsilon
N^{1/\bar{\nu}};a,b)$ with $\gamma=1$. The scaling function
$\tilde{\phi}(x;a,b)$ is defined in Eq.~(\ref{sus_s_f}).

\subsection{Scale-free networks}\label{sec:III.B}

Consider the SF degree distribution $P(k)\sim k^{-\lambda}$ for
$k_0 \le k \le k_c$ and $0$ otherwise with the upper cutoff
$k_c\sim N^{1/\omega}$ and the lower cutoff $k_0=\mathcal{O} (1)$.
We are interested in the cutoff exponent $\omega\ge \omega_{\rm nat}=\lambda-1$
as considered in~\cite{CP-S2007,Castellano2007,Boguna2009,JDNoh2009}
where the cutoff-dependent FSS in the CP
on annealed SF networks.
In general, the expansion of Eq.~(\ref{s_c}) for small $u_0$ and $\tilde{h}$
is singular as $z_n$ diverges in the $N\rightarrow\infty$ limit for
$n\ge \lambda-1$.  So one should treat the nonanalyticity carefully.
Furthermore, there is a power-law finite-size correction in the critical
inverse temperature $\tilde{K}_c$, which plays
an intricate role in the critical FSS.
For all $\lambda>3$,
the average magnetization is again $m\simeq a \langle\tilde{m}\rangle$ and
the magnetic susceptibility is identical to that in the exponential networks.

\subsubsection{Finite-size behavior of $z_n$}

As a degree $k$ is an integer, the standard precise expression for
the degree distribution is
\begin{equation}\label{p_k}
P(k)=ck^{-\lambda}\sum_{j=k_0}^{k_c} \delta_{k,j},
\end{equation}
where  the normalization factor $c$ is given by
$c^{-1}=\sum_{j=k_0}^{k_c} j^{-\lambda}$ with $k_c=dN^{1/\omega}$.
Then, the degree moments $z_n$ are given by
\begin{equation}
z_n=c\sum_{j=k_0}^{k_c} j^{-\lambda+n}.
\end{equation}

For large $k_c$~(large $N$), we have finite-size corrections
for the normalization factor as
\begin{equation}
c^{-1}\simeq c^{-1}_\infty -k_c^{-(\lambda-1)}/(\lambda-1),
\end{equation}
with $c^{-1}_\infty=\zeta(\lambda,k_0)$
where the Hurwitz zeta function is defined as
$\zeta(s,l)\equiv \sum_{j=0}^\infty (j+l)^{-s}$~\cite{Elizalde1994}.

Similarly, we have, up to the leading order in $k_c$,
\begin{equation}
z_n\simeq
\left\{\begin{array}{lll}
\displaystyle  z_n^\infty -c_\infty
\frac{k_c^{-[\lambda-(n+1)]}}{\lambda-(n+1)} \ \ &\mbox{for} \  n<\lambda-1 & \\
\displaystyle  c_\infty \ln k_c &\mbox{for} \  n=\lambda-1 & \\
\displaystyle  c_\infty \frac{k_c^{(n+1)-\lambda}}{(n+1)-\lambda}
  &\mbox{for} \  n>\lambda-1 &,
\end{array}\right.
\end{equation}
with $z_n^\infty=c_\infty\zeta(\lambda-n,k_0)=
\zeta(\lambda-n,k_0)/\zeta(\lambda,k_0)$.

The critical parameter $\tilde{K}_c=z_1/z_2$ also has a finite-size
correction as
\begin{equation}\label{cpfss}
\tilde{K}_c\simeq \tilde{K}_c^\infty \left[1+e N^{-\alpha}\right],
\end{equation}
with
$$\tilde{K}_c^\infty=z_1^\infty/z_2^\infty=\zeta(\lambda-1,k_0)/
\zeta(\lambda-2,k_0),$$
\begin{equation}\label{def_e}
e=d^{-(\lambda-3)}/[(\lambda-3)\zeta(\lambda-2,k_0)] ,
\end{equation}
and
\begin{equation}\label{def_alpha}
\alpha = (\lambda-3)/\omega .
\end{equation}

\subsubsection{$\lambda>5$}

For $\lambda>5$, $z_n$ is finite up to $n=4$. Hence, the expansion of
$\tilde{m}$ and $f(\tilde{m})$ are the same as
those in the exponential networks
up to the order of $u_0^3$ and up to the order of $\tilde{m}^4$, respectively,
as in Eqs.~(\ref{m_u0_ll}) and (\ref{freg}).
Therefore, their critical behaviors are identical to those in the
exponential networks, in terms of the parameters $a$, $b$, $\epsilon$, and $N$.

\begin{figure}[b]
\includegraphics[width=0.9\columnwidth]{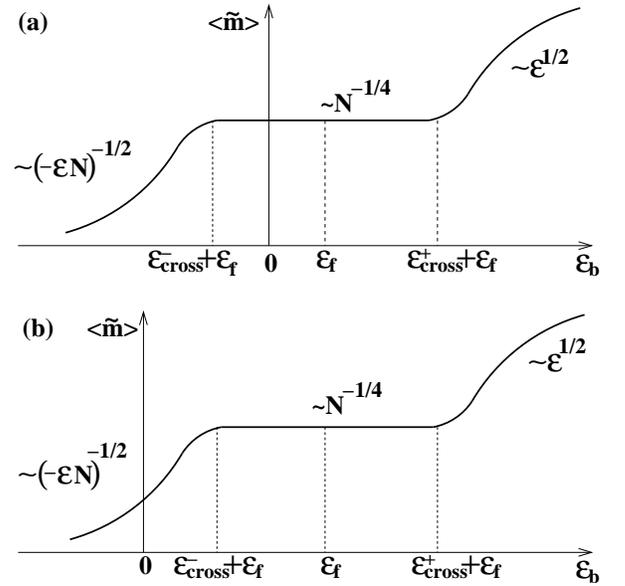}
\caption{Schematic plot of $\langle\tilde{m}\rangle$ versus $\epsilon_b$
in the annealed SF networks with $\lambda>5$ and (a) $\omega_{\rm nat}\le
\omega < 2(\lambda-3)$ and
(b) $\omega>2(\lambda-3)$. Note that the bulk critical point $\epsilon_b=0$ is
outside of the critical region in (b).}
\label{fig2}
\end{figure}

However, unlike the exponential networks, $\epsilon =
(\tilde{K}-\tilde{K}_c) / \tilde{K}_c$ has a power-law
finite-size correction due to the $N$-dependence of $\tilde{K}_c$. From
Eq.~(\ref{cpfss}), one finds that
\begin{equation}\label{kc_corr}
\epsilon\simeq\epsilon_b -\epsilon_f(N)=\epsilon_b-e N^{-\alpha},
\end{equation}
where $\epsilon_b\equiv (\tilde{K}-\tilde{K}_c^\infty)/\tilde{K}_c^\infty$
is a deviation from the bulk critical temperature and
$\alpha=(\lambda-3)/\omega$.
Therefore, the FSS form is given in terms of $\epsilon_b$ as
\begin{equation}
\label{fss-sf}
\langle\tilde{m}(\epsilon_b,N)\rangle=N^{-\beta/\bar\nu} \tilde{\psi}
[(\epsilon_b-\epsilon_f) N^{1/\bar\nu}],
\end{equation}
with $\beta=1/2$ and $\bar\nu=2$, which shows a simple horizontal shift
of the order parameter curve in Fig.~\ref{fig1} to the right~(see
Fig.~\ref{fig2}).

The order parameter follows the same scaling law of Eq.~(\ref{exp:m_tc})
at the $N$-dependent pseudo critical temperature at $\epsilon=0$ or
$\epsilon_b=\epsilon_f$.
On the other hand, at the bulk critical temperature at $\epsilon_b=0$,
the order parameter is given by $\langle \tilde{m} \rangle_{b,c}
 = N^{-\beta/\bar{\nu}} \tilde{\psi}(-eN^{-\alpha+1/\bar{\nu}})$.

For $\alpha>1/2$~[$\omega_{\rm nat}\le \omega < 2(\lambda-3)$],
the correction $\epsilon_f$ is not big enough to shift $\epsilon_{b,{\rm
cross}}^- \equiv \epsilon_{\rm cross}^- + \epsilon_f $ with $\epsilon_{\rm
cross}^-$ in Eq.~(\ref{e_cross_-})
to cross the bulk critical point $\epsilon_b=0$~[Fig.~\ref{fig2}(a)].
Therefore, there is no characteristic change in the critical scaling
by this shift, except the appearance of a higher-order correction to
scaling like $\mathcal{O}~(N^{-(\alpha-1/2)})$.

For $\alpha<1/2$~[$\omega>2(\lambda-3)$], $\epsilon_{b,{\rm cross}}^-$
becomes positive and both crossovers take place in the side of
$\epsilon_b>0$~[Fig.~\ref{fig2}(b)].
The bulk critical point is now in the region left to
$\epsilon_{b,{\rm cross}}^-$, where the scaling function behaves as
$\tilde\psi(x)\simeq \sqrt{2/(\pi a)}\ (-x)^{-1/2}$.
At $\alpha=1/2$~[$\omega=2(\lambda-3)$],
the scaling variable is finite~($x=-\epsilon_f N^{1/\bar{\nu}} = -e$).

Therefore, we have the critical FSS at $\epsilon_b=0$ as
\begin{equation}\label{exp:m_b_tc}
\langle \tilde{m}\rangle_{b,c} = \left\{
 \begin{array}{lcc}
   A_e N^{-1/4} & & (\alpha>1/2)  \\[2mm]
   \tilde{A}_e N^{-1/4} & & (\alpha = 1/2)  \\[2mm]
   B_e N^{-(1-\alpha)/2} & & (\alpha<1/2) ,
 \end{array}\right.
\end{equation}
where $\tilde{A}_e = \tilde{\psi}(-e)$ and $B_e = [2/(\pi a e)]^{1/2}$ with
$e$ in Eq.~(\ref{def_e}).

\begin{figure}[t]
\includegraphics[width=0.9\columnwidth]{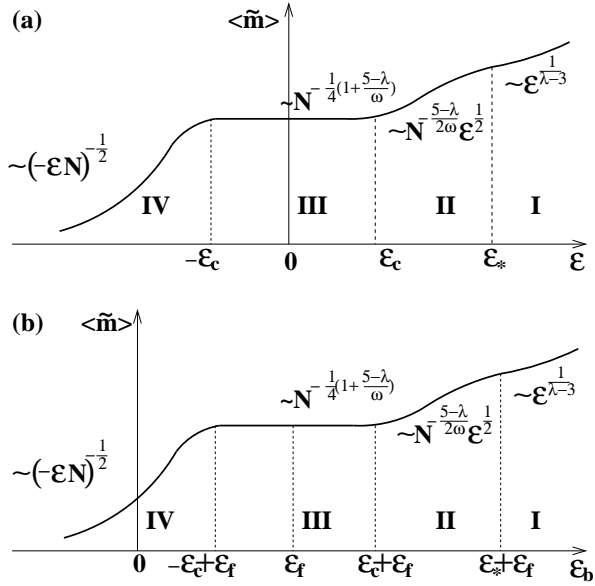}
\caption{Schematic plots of $\langle\tilde{m}\rangle$ versus (a) $\epsilon$ and
$\langle\tilde{m}\rangle$ versus (b) $\epsilon_b$ in the annealed SF networks
with $3<\lambda<5$ and $\omega>\omega_{\rm nat}$.}
\label{fig3}
\end{figure}

\subsubsection{$3 < \lambda < 5$}

For $3 < \lambda < 5$, $z_1$ and $z_2$ are finite, but $z_4$ diverges as $z_4\sim k_c^{5-\lambda}\sim N^{(5-\lambda)/\omega}$ as well as $b= z_4(z_1/z_2)^4$.
In the thermodynamic limit or for $u_0k_c +\tilde{h}\gg 1$ in finite size networks,
Eq.~(\ref{s_c}) has the singular expansion as
\begin{equation}\label{m_u0_sl_2}
\tilde{m} = \tilde{h} +\frac{z_2}{z_1} u_0 - \frac{q}{z_1} u_0^{\lambda-2}+ \mathcal{O}(\tilde{h}^2),
\end{equation}
with a constant $q\simeq c\int dx x^{1-\lambda}(x-{\rm tanh}x)>0$.
For $u_0k_c +\tilde{h}\ll 1$ in finite networks, the series expansion becomes regular as
\begin{equation}\label{m_u0_sl_1}
\tilde{m} = \tilde{h}+ \frac{z_2}{z_1}u_0 - \frac{z_4}{3z_1} u_0^3 + \mathcal{O}(u_0^5,\tilde{h}^2) \ .
\end{equation}

Then, the free-energy density in finite networks is given by
\begin{equation}\label{free_s}
f(\tilde{m}) = -\ln{2} -a\tilde{h}\tilde{m} -\frac{a}{2}\epsilon\tilde{m}^2 + Q(\tilde{m}) ,
\end{equation}
where
\begin{equation}
Q(\tilde{m}) = \left\{ \begin{array}{ccc}
\displaystyle\frac{b}{12}\ \tilde{m}^4 &\mbox{ for }
\tilde{m} \ll \left({z_2}/{z_1}\right){k_c}^{-1}+\tilde{h}& \\
[4mm]
\displaystyle p \frac{z_1^{\lambda-1}}{ z_2^{\lambda-1}}\ \tilde{m}^{\lambda-1}
&\mbox{ for } \tilde{m} \gg \left({z_2}/{z_1}\right){k_c}^{-1}+\tilde{h} ,&
\end{array} \right.
\label{q_m}
\end{equation}
with a constant $p\simeq c\int dx x^{-\lambda}(x^2/2-\ln\cosh x)>0$.
Note that $b=(z_1/z_2)^4 z_4\simeq b_0 N^{(5-\lambda)/\omega}$ with
\begin{equation}\label{b_0}
b_0=c_\infty (z_1^\infty/z_2^\infty)^4 d^{5-\lambda}/(5-\lambda) .
\end{equation}

In the thermodynamic limit, $k_c$ becomes infinite and the free energy density
expansion is singular for all $\tilde{m}>0$. At $\tilde{h}=0$, the order parameter scales
for $\epsilon>0$ as
\begin{equation}\label{m_bulk}
\langle\tilde{m}\rangle \simeq  C \epsilon^{\beta} ,
\end{equation}
with
\begin{equation}\label{beta-sf}
\beta = 1/(\lambda-3),
\end{equation}
and
\begin{equation}\label{C-sf}
C=(z_2^\infty/z_1^\infty)\{z_2^\infty/[p(\lambda-1)]\}^{1/(\lambda-3)} .
\end{equation}

It is straightforward to calculate the FSS at $\epsilon=0$
by performing the integral in Eq.~(\ref{finite}) using Eqs.~(\ref{free_s}) and (\ref{q_m}).
For $\omega>\omega_{\rm nat}$, the integral in the region of $\tilde{m} \ll \left({z_2}/{z_1}\right){k_c}^{-1}$
dominates and we find
\begin{equation}\label{m_critical}
\langle\tilde{m}\rangle_c\simeq A_s N^{-[1+(5-\lambda)/\omega]/4}\sim
(bN)^{-1/4} ,
\end{equation}
with
\begin{equation}\label{A_s}
A_s=(12/b_0)^{1/4}\Gamma(1/2)/\Gamma(1/4) .
\end{equation}
At $\omega=\omega_{\rm nat}$, the integrals in both regions contribute,
but the critical FSS does not change except its amplitude.

For $\epsilon<0$, we have $\langle\tilde{m}\rangle\simeq \sqrt{2/(\pi a)}\ (-\epsilon)^{-1/2} N^{-1/2}$.
So, the crossover occurs at $\epsilon_{{\rm cross}}^-\simeq
-[2/(\pi aA_s^2)]N^{-1/\bar\nu_-}$ with $\bar\nu_-=2/[1-(5-\lambda)/\omega]$.

At small positive values of $\epsilon$,
we have a nonzero solution for $\tilde{m}$ in the region of $\tilde{m}\ll (z_2/z_1)k_c^{-1}$
as
\begin{equation}\label{m_r2}
\langle\tilde{m}\rangle\simeq \sqrt{3a/b_0}\ N^{-(5-\lambda)/(2\omega)}
\epsilon^{1/2} \sim (\epsilon/b)^{1/2},
\end{equation}
where $\epsilon\ll f_1 N^{-(\lambda-3)/\omega}$ with
$f_1=c_\infty d^{-(\lambda-3)}/[3z_2^\infty(5-\lambda)]$.
For larger $\epsilon$, we have the bulk solution, Eq.~(\ref{m_bulk}),
for $\tilde{m}$ in
the region of $\tilde{m}\gg (z_2/z_1)k_c^{-1}$, where
$\epsilon\gg f_2 N^{-(\lambda-3)/\omega}$ with $f_2=p(\lambda-1) d^{-(\lambda-3)}/z_2^\infty$.

Hence, there are two crossovers in the side of $\epsilon>0$. The first crossover occurs at
$$\epsilon_{{\rm cross}}^{+,1}\simeq [b_0A_s^2/(3a)]N^{-1/\bar\nu_{+,1}} ,$$
with $\bar\nu_{+,1}=2/[1-(5-\lambda)/\omega],$
which is the same as $\bar\nu_{-}$. Then second crossover occurs at
$$\epsilon_{{\rm cross}}^{+,2}\simeq [b_0C^2/(3a)]^{-(\lambda-3)/(5-\lambda)} N^{-1/\bar\nu_{+,2}} ,$$
with $\bar\nu_{+,2}=\omega/(\lambda-3)$.
Note that $1/\bar\nu_{+,2}$
coincides incidentally  with $\alpha$ in Eq.~(\ref{kc_corr}).
For convenience, we denote $\epsilon_{{\rm cross}}^{+,1}$ by $\epsilon_c$,
$\epsilon_{{\rm cross}}^{+,2}$ by $\epsilon_*$, $\bar\nu_{+,1}=\bar\nu_-$ by $\bar\nu_c$,
and $\bar\nu_{+,2}$ by $\bar\nu_*$. They are summarized as
\begin{eqnarray}
\bar{\nu}_c &=& 2/[1-(5-\lambda)/\omega],  \\
\bar{\nu}_* &=& \omega/(\lambda-3) .
\end{eqnarray}

The order parameter is plotted against $\epsilon$ in Fig.~\ref{fig3}(a), where we have
one more distinct scaling regime compared to the case for $\lambda>5$.
In (bulk) regime I~($\epsilon> \epsilon_*$), the bulk scaling is valid where the system is free from
any finite size effect. In (intermediate) regime II~($\epsilon_c< \epsilon< \epsilon_*$), the system behaves
as in a SF network with infinite $N$ but with finite $k_c$. In (critical) regime III
($|\epsilon|< \epsilon_c$), the system feels both finite $N$ and finite $k_c$.
Finally, the ordinary scaling in the disordered phase appears in (disordered) regime IV,
where only finite $N$ matters.

Summing up the results, we need two different scaling functions, $\tilde\psi_c$ and
$\tilde\psi_*$, describing
the critical region and the crossover region to the bulk regime, respectively,
for the forced cutoff~($\omega>\omega_{\rm nat}$).
First, near $\epsilon\approx 0$, we have
\begin{equation}
\label{fss-sf-1}
\langle\tilde{m}(\epsilon,N)\rangle =N^{-\tilde\beta/\bar\nu_c} \ \tilde{\psi}_c (\epsilon N^{1/\bar\nu_c}),
\end{equation}
with $\tilde\beta=(\omega+5-\lambda)/[2(\omega-5+\lambda)]$.
The scaling function $\tilde\psi_c(0)\simeq A_s$~[Eq.~(\ref{A_s})],
$\tilde\psi_c(x)\simeq \sqrt{2/(\pi a)}\ (-x)^{-1/2}$
for $x\rightarrow -\infty$, and $\tilde\psi_c(x)\simeq \sqrt{3a/b_0}\
x^{1/2}$~[Eq.~(\ref{b_0})] for $x\rightarrow \infty$.
Due to the crossover to the bulk regime, this scaling function
is valid only up to $x \sim N^{[1-(\lambda-1)/\omega]/2}$, which diverges with $N$.

Second, near $\epsilon\approx \epsilon_*$, we have
\begin{equation}
\label{fss-sf-2}
\langle\tilde{m}(\epsilon,N)\rangle=N^{-\beta/\bar\nu_*} \ \tilde{\psi}_* (\epsilon N^{1/\bar\nu_*}),
\end{equation}
with $\beta=1/(\lambda-3)$.
The scaling function $\tilde\psi_*(x)\simeq C x^{\beta}$~[Eq.~(\ref{C-sf})]
for $x\rightarrow \infty$ and
$\tilde\psi_*(x)\simeq \sqrt{3a/b_0}\ x^{1/2}$ for small $x$, but larger than
$\sim N^{-[1-(\lambda-1)/\omega]/2}$, which vanishes as $N\rightarrow\infty$.
At $\omega=\omega_{\rm nat}$, the intermediate regime II vanishes as $\epsilon_c\sim \epsilon_*$
so that the two scaling functions merge into a single scaling function
with $\tilde\beta=\beta=1/(\lambda-3)$ and $\bar\nu_c=\bar\nu_*=(\lambda-1)/(\lambda-3)$.

Now, in terms of the bulk parameter $\epsilon_b$, we need to replace $\epsilon$ by $\epsilon_b-\epsilon_f$
in all scaling equations. It implies the simple horizontal shift of the
order parameter curve of Fig.~\ref{fig3} to the right by the amount of
$\epsilon_f=eN^{-\alpha}$ with $\alpha = (\lambda-3)/\omega$~[Fig.~\ref{fig3}(b)].
For any $\omega \ge \omega_{\rm nat}$, we find that
$\epsilon_{b,{\rm cross}}=\epsilon_{{\rm cross}}^-+\epsilon_f$
becomes always positive. Therefore, the critical FSS at $\epsilon_b=0$ is
$\langle\tilde{m}\rangle_{b,c}=N^{-\tilde\beta/\bar\nu_c}\
\tilde\psi_c(-\epsilon_f N^{1/\bar\nu_c})$,
which results in
\begin{equation}\label{m_bulk_critical}
\langle\tilde{m}\rangle_{b,c} \simeq B_e N^{-[1-(\lambda-3)/\omega]/2} ,
\end{equation}
with $B_e = [2/(\pi ae)]^{1/2}$. At $\omega=\omega_{\rm nat}$,
the critical scaling does not change except for its amplitude as
$\langle\tilde{m}\rangle_{b,c}\sim N^{-1/(\lambda-1)}$.

\subsection{Comparison to quenched networks}

In quenched networks, it is difficult to derive analytically the FSS
for any model due to the presence of quenched disorder.
Even in the case that quenched disorder fluctuations are negligible,
quenched links generate the finite correlations in neighboring nodes,
which are responsible for the critical point shift~(mass shift)
by a finite amount.
This mass renormalization process should involve the finite-size
correction which determines the FSS of the ({\em pseudo-})
critical point and the FSS of the order parameter follows.

Recently, Hong {\em et al.}~\cite{HHong2007} conjectured
the FSS exponent based on the droplet-excitation (hyperscaling)
argument and phenomenological theory. They also numerically
confirmed that the FSS exponent for the Ising model
is $\bar\nu=2$ for quenched exponential networks
as well as for the quenched SF networks with $\lambda>5$.
For $3<\lambda<5$, $\bar\nu=(\lambda-1)/(\lambda-3)$,
regardless of the cutoff exponent $\omega$
if it is not too strong~($\omega<\lambda$).

For exponential networks,
we find the same FSS for annealed and quenched networks.
The annealed SF networks with $\lambda>5$ exhibit essentially
the same FSS as the quenched SF networks,
but the additional finite-size correction
on the critical point generates a different FSS
on the order parameter at the bulk critical point
for $\omega>2(\lambda-3)$.
For $\omega_{\rm nat}\le \omega < 2(\lambda-3)$,
this additional correction is irrelevant.

The annealed SF networks with $3<\lambda<5$ exhibit the anomalous
FSS characterized by the combination of two different single-parameter
scaling functions~(or two-parameter scaling)
with the anomalous intermediate regime for any $\omega>\omega_{\rm nat}$,
which is generically distinct from the quenched SF networks.
However, at $\omega=\omega_{\rm nat}$,
the intermediate regime disappears and the FSS can be described
by the ordinary single-parameter scaling function
with the same exponent $\bar\nu=(\lambda-1)/(\lambda-3)$
as in the quenched SF networks.

\section{Numerical Results}\label{sec4}

\begin{figure}[]
\begin{tabular}{c}
\includegraphics[width=\columnwidth]{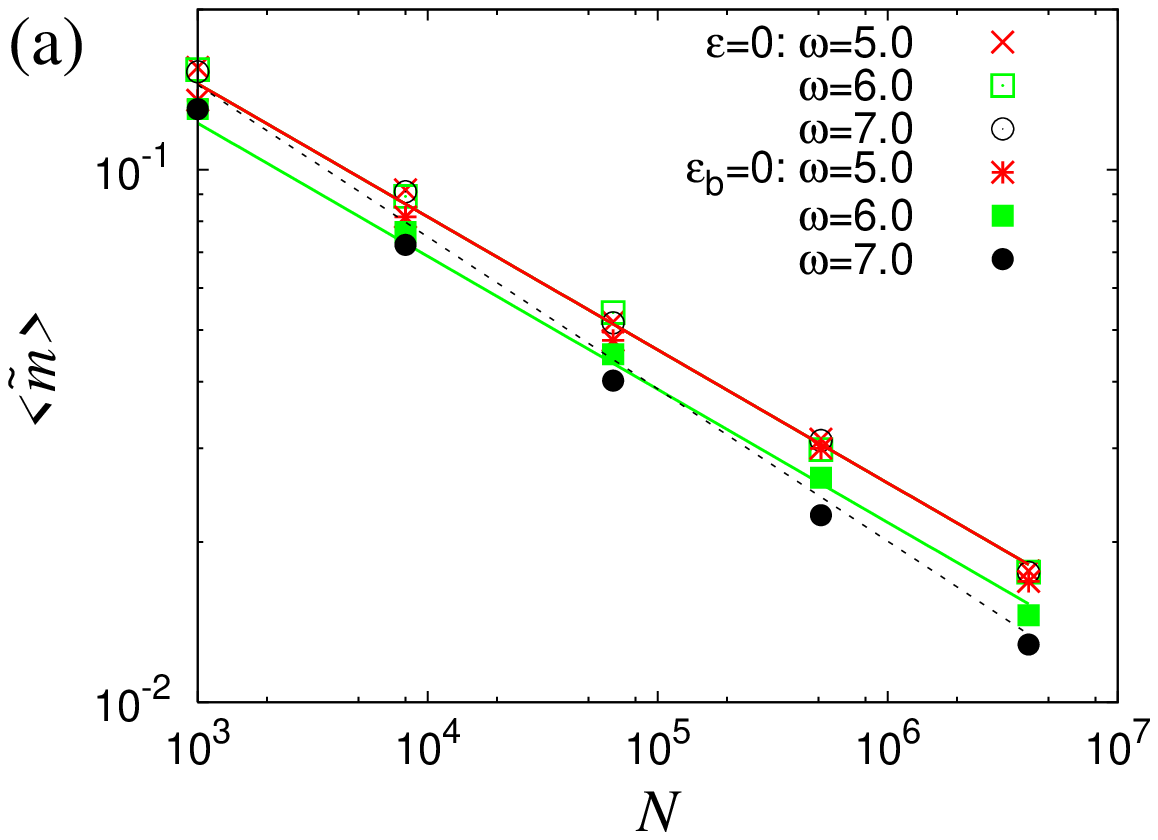} \\
\includegraphics[width=\columnwidth]{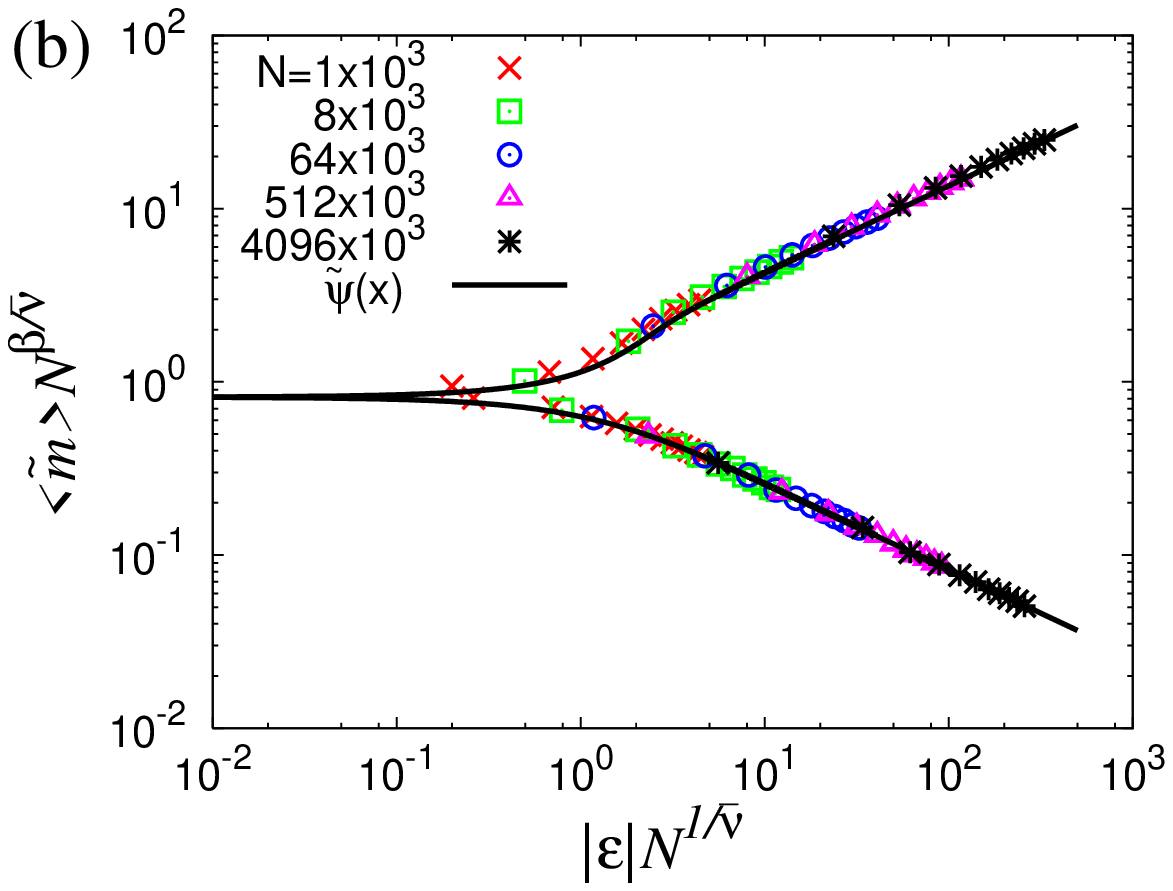}
\end{tabular}
\caption{(Color online) Monte Carlo simulation data at $\lambda=6$.
(a) The order parameters $\langle\tilde{m}\rangle_c$ at
$\epsilon=0$ and $\langle\tilde{m}\rangle_{b,c}$ at $\epsilon_b=0$
with different values of $(d,\omega)=(2.25,5)$, $(3,6)$, and $(3,7)$
are plotted with symbols.
They are compared with the analytic results of Eqs.~(\ref{exp:m_tc}) and
(\ref{exp:m_b_tc}) which are drawn with lines. Numerical values of the
coefficients are $A_e \simeq 0.815~860$,
$\tilde{A}_e \simeq 0.687~983$, and $B_e\simeq 1.039~20$.
(b) Scaling plot of $\langle \tilde{m}\rangle N^{\beta/\bar{\nu}}$ versus
$|\epsilon| N^{1/\bar{\nu}}$ at $\omega=6$ with $\beta=1/2$ and
$\bar{\nu}=2$. Data with different values
of $N$ fall onto the scaling function $\tilde{\psi}(|\epsilon|N^{1/2})$
drawn with the solid curve.}
\label{fig4}
\end{figure}

\begin{figure*}[th]
\begin{tabular}{cc}
\includegraphics[width=\columnwidth]{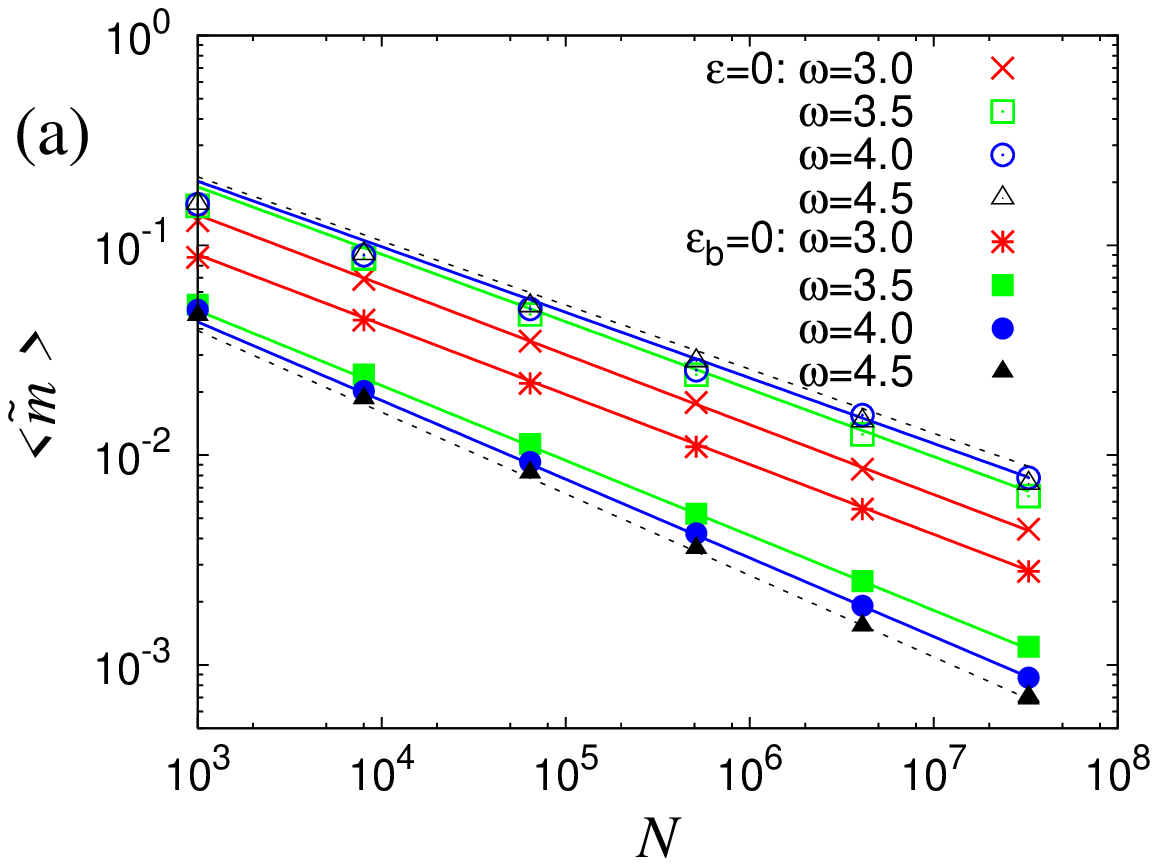}&
\includegraphics[width=\columnwidth]{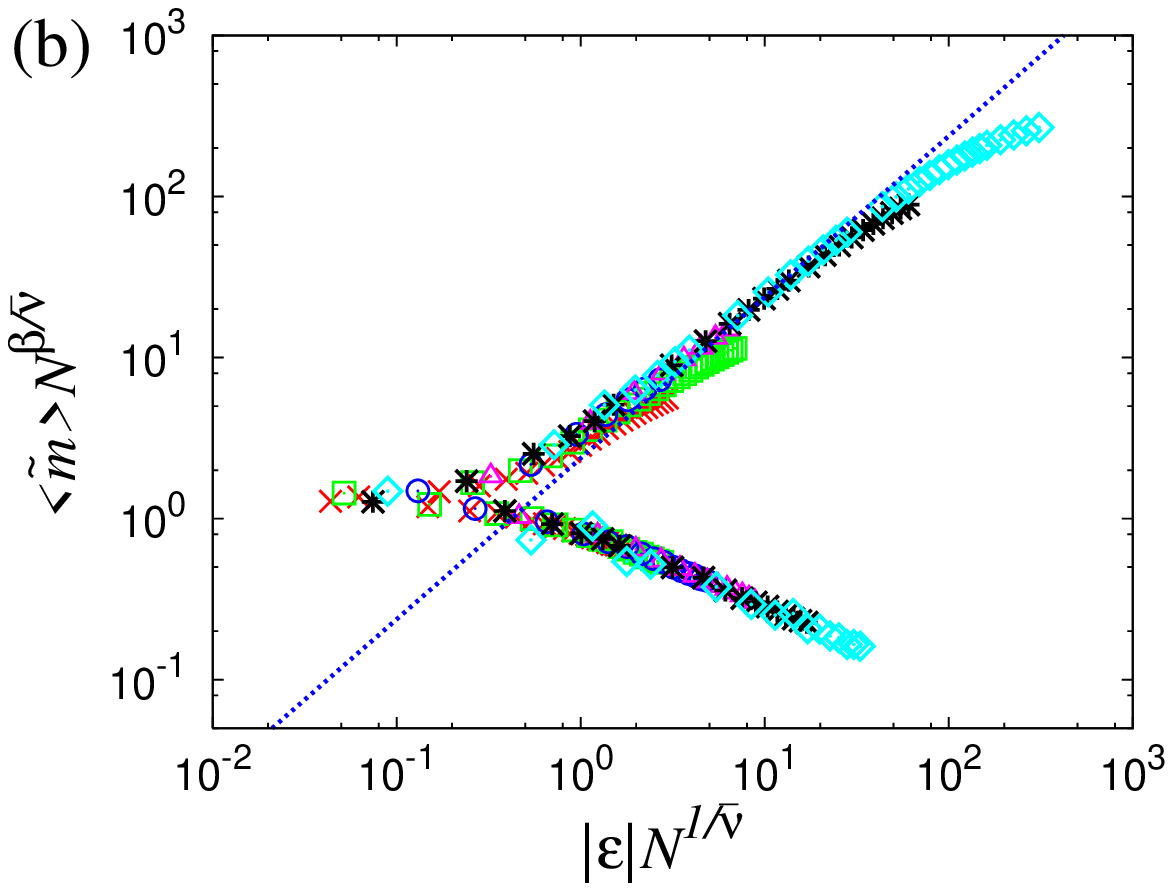}\\
\includegraphics[width=\columnwidth]{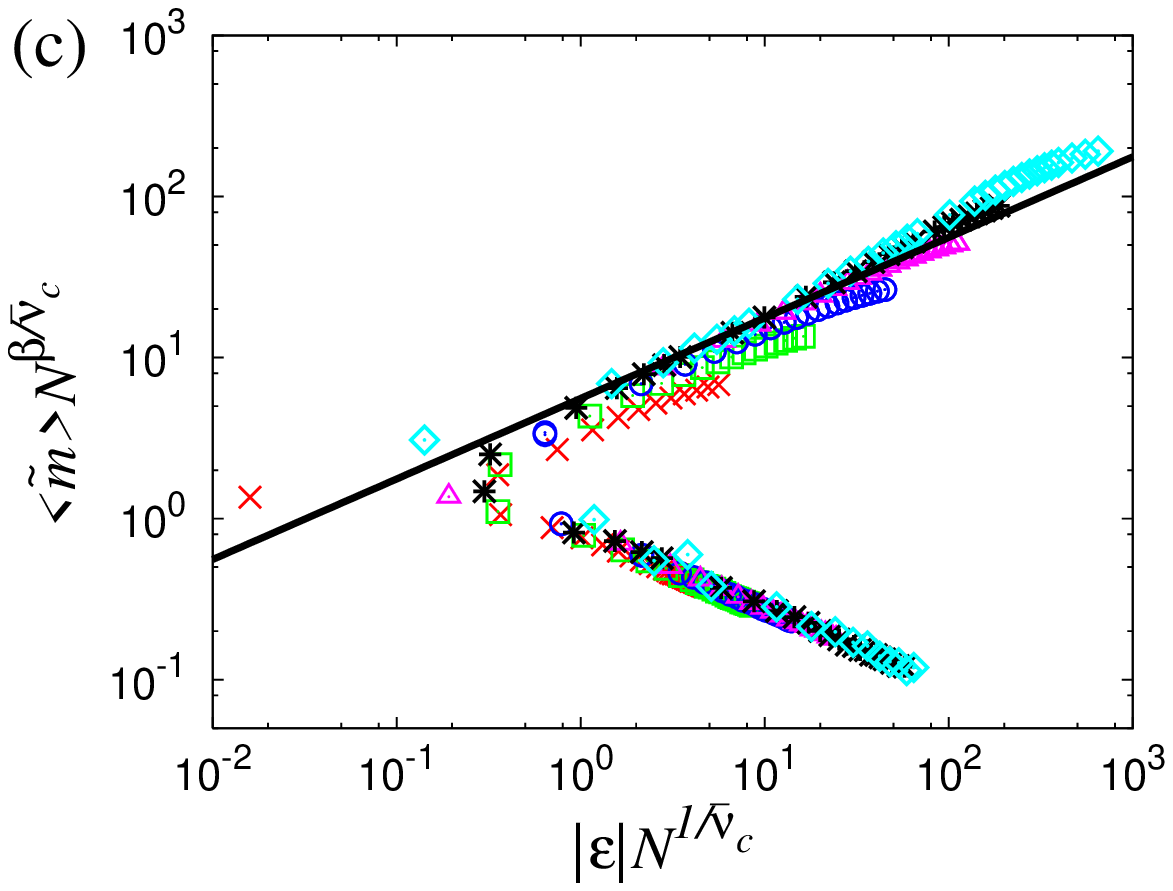}&
\includegraphics[width=\columnwidth]{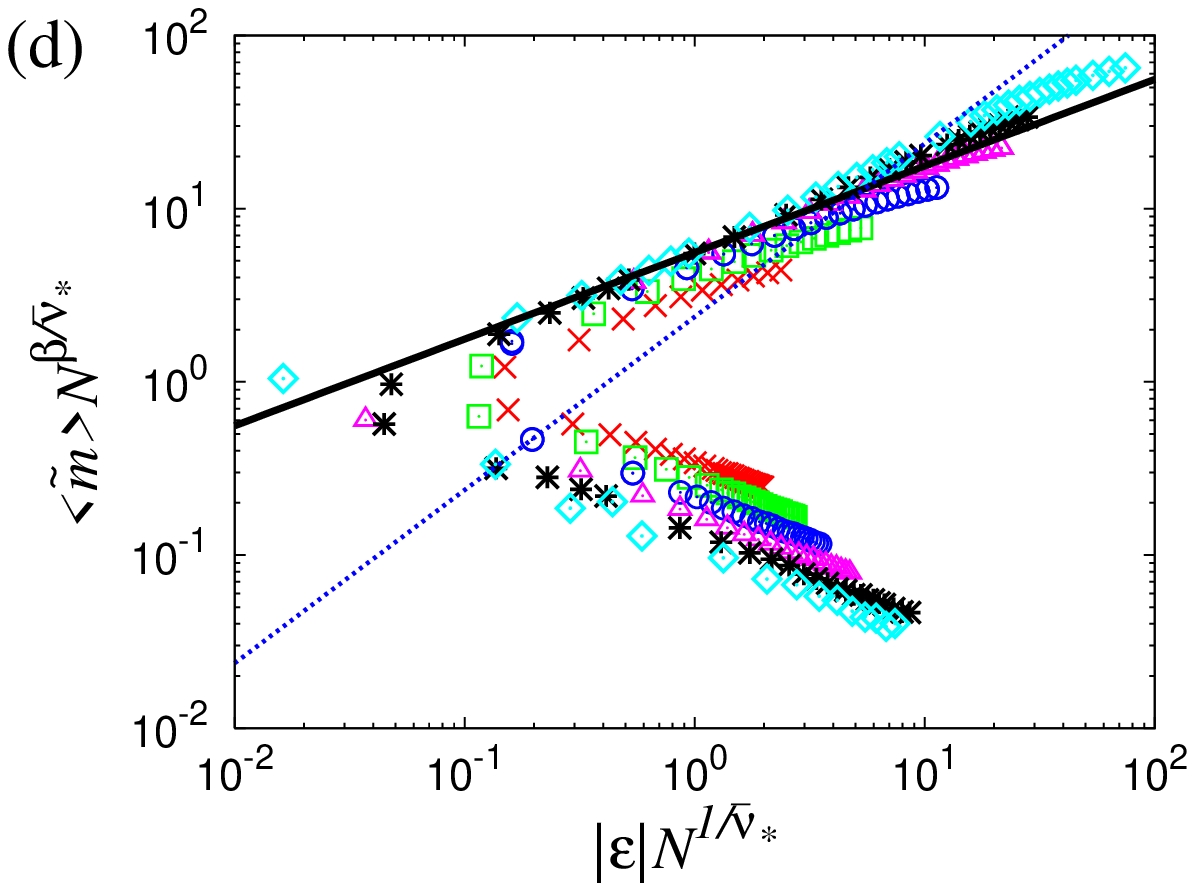}
\end{tabular}
\caption{(Color online) Monte Carlo simulation data at $\lambda=4$.
The parameter values of $d$ are $2.57$ for $\omega=3$
and $1$ for the other values of $\omega$.
(a) Numerical data for $\langle\tilde{m}\rangle_c$ and
$\langle\tilde{m}\rangle_{b,c}$ represented with symbols
are compared with the analytic results represented with straight lines.
(b) Scaling plot of $\langle\tilde{m}\rangle N^{1/(\lambda-1)}$ versus
$|\epsilon|N^{(\lambda-3)/(\lambda-1)}$ at $\omega=\omega_{\rm nat}=3$. The
straight line has the slope of $\beta=1/(\lambda-3)=1$. We also show
the scaling collapse of the numerical data at $\omega=4$ according to the
FSS forms in (c) Eq.~(\ref{fss-sf-1}) and (d) Eq.~(\ref{fss-sf-2}).
The solid line and the dashed line have a slope of $1/2$ and $\beta=1/(\lambda-3)=1$,
respectively. We use same symbols of Fig.~\ref{fig4}(b)
and the symbol $\diamond$ for $N=32~768\times 10^3$.}
\label{fig5}
\end{figure*}

We performed extensive Monte Carlo (MC) simulations
in the annealed SF networks at various values of $\lambda$ and $\omega$
to confirm the analytic results
in Sec.~\ref{sec:III.B}. Especially we focus our attention
on the cutoff dependent FSS behavior.
In practice,
we consider the Ising model on the fully connected network
with the heterogeneous coupling constants given by the Hamiltonian in Eq.~(\ref{H_s})
at all $\tilde{h}_i=0$ and set $J=k_B=1$.

Using the standard Metropolis single spin update rule,
we run MC simulations up to $2\times 10^3-10^4$ MC steps
for system sizes up to $N = 32~768\times 10^3$. The MC data are averaged over
100 independent samples of initial spin configurations as well as the thermal (temporal) average
after discarding the data up to $10^3$ MC steps for the equilibration.

First, we need to choose a degree sequence $\{k_i\}=\{k_1,\ldots,k_N\}$
in accordance with a given degree distribution $P(k) = c k^{-\lambda}$ for $k_0\le k \le k_c$
as in Eq.~(\ref{p_k}), with $k_0=\mathcal{O}(1)$ and
$k_c={\rm int}[dN^{1/\omega}]$, where
${\rm int}[x]$ denotes the integer part of $x$.
Let $N_k$ be the number of nodes with degree $k$.
Such a degree sequence can be generated
{\em deterministically}~\cite{JDNoh2009} by applying the rule
\begin{equation}\label{det_way}
\sum_{k'=k}^{k_c} N_{k'}={\rm int}\left[N\sum_{k'=k}^{k_c} P(k')\right]
\end{equation}
to $N_k$ for all $k$ in the descending order from $k=k_c$.

The maximum degree $k'_c$ thus obtained may be different form the target
value $k_c$.
In fact, $k'_c$ can be estimated from the
condition $N\sum_{k=k'_c}^{d N^{1/\omega}} P(k) = 1$, which yields
$k'_c = d N^{1/\omega} [1 + \mathcal{O}(N^{-1+\omega_{\rm nat}/\omega}) ]$
for $\omega > \omega_{\rm nat}$. Therefore, Eq.~(\ref{det_way}) indeed yields
the degree cutoff scaling with the prescribed values of $d$ and $\omega$ only with a
higher-order correction. However, when $\omega=\omega_{\rm nat}$, we
find that $k'_c = d' N^{1/\omega}$ with
\begin{equation}\label{d_prime}
d' = d ( 1 + (\lambda-1) d^{\lambda-1} \zeta(\lambda,k_0))^{-1/(\lambda-1)}.
\end{equation}
When one compares numerical data with the analytic results, the modified value $d'$
should be used for $\omega=\omega_{\rm nat} = \lambda-1$.
In this section, we use the degree sequences generated
deterministically from Eq.~(\ref{det_way}) for various $N$, $\lambda$, $d$,
and $\omega$ with fixed $k_0=3$.

Monte Carlo simulation data for $\lambda=6$ are presented in
Fig.~\ref{fig4}. We first test whether the magnetizations $\langle
\tilde{m}\rangle_c$ at the pseudo critical temperature with $\epsilon=0$
and $\langle \tilde{m}\rangle_{b,c}$ at the bulk critical temperature
with $\epsilon_b=0$ scale as in
Eqs.~(\ref{exp:m_tc}) and (\ref{exp:m_b_tc}), respectively.
In order to cover the three cases of Eq.~(\ref{exp:m_b_tc}), we choose
$\omega = 5$, $6$, and $7$, which correspond to $\alpha = 3/5$, $1/2$, and
$3/7$, respectively. These numerical data in Fig.~\ref{fig4}(a)
are in good agreement with the analytic results.

Our analytic theory predicts the full shape of the scaling function
as well as the scaling exponents. We examine validity of the FSS form
in Eq.~(\ref{m_fss_form}) in Fig.~\ref{fig4}(b).
We present the scaling plot of $\langle \tilde{m}(\epsilon,N)\rangle$
against $|\epsilon| N^{1/\bar{\nu}}$ using the Monte Carlo data with
$\lambda=6$ and $\omega=6$. These data match perfectly well with the
analytic curve for the scaling function $\tilde{\psi}(|\epsilon| N^{1/2})$
in Eq.~(\ref{m_s_f}).

We proceed to the case with $\lambda=4$, where the FSS behavior is more
complicated. We first examine the FSS of $\langle\tilde{m}\rangle_c$ at
$\epsilon=0$ and $\langle\tilde{m}\rangle_{b,c}$ at $\epsilon_b=0$.
They are predicted to follow the power law given in
Eqs.~(\ref{m_critical}) and (\ref{m_bulk_critical}), respectively, when
$\omega>\omega_{\rm nat}$. When $\omega=\omega_{\rm nat}$, the scaling
is given by the same power law but with modified amplitudes.
Figure~\ref{fig5}(a) presents the plots of $\langle\tilde{m}\rangle_c$ and
$\langle\tilde{m}\rangle_{b,c}$ against $N$ at $\omega=3$, $3.5$, $4$, and $4.5$,
which agree well with the theoretical curves.

When $\omega=\omega_{\rm nat}$, the FSS is governed with the single scaling
variable $\epsilon N^{(\lambda-3)/(\lambda-1)}$. In Fig.~\ref{fig5}(b),
we present the scaling plot of $\langle \tilde{m}\rangle N^{1/(\lambda-1)}$
against $|\epsilon| N^{(\lambda-3)/(\lambda-1)}$ at $\lambda=4$ and
$\omega=3$. A good data collapse supports that FSS form with the single
scaling variable. We note that the bulk scaling behavior
$\langle\tilde{m}\rangle \sim \epsilon^{\beta}$ with $\beta=1/(\lambda-3)$
sets in only for $N \gg 10^6$.

We also examine the FSS behavior at $\omega=4~(>\omega_{\rm nat})$. Here, the
FSS is governed with two scaling variables
$\epsilon N^{1/\bar{\nu}_c}$ and $\epsilon N^{1/\bar{\nu}_*}$. Hence, one
cannot expect a data collapse over the whole regions in a scaling plot.
We first test the scaling form of Eq.~(\ref{fss-sf-1}), which is valid in
regimes II, III, and IV. The scaling plot of $\langle\tilde{m}\rangle
N^{\tilde{\beta}/\bar{\nu}_c}$ against the scaling variable
$|\epsilon|N^{1/\bar{\nu}_c}$ is presented in Fig.~\ref{fig5}(c). We observe
a reasonably good data collapse in regimes II, III, and IV except for
small network sizes. In the $\epsilon>0$ side, the numerical data
align along a straight line of slope $1/2$, which reflects the scaling
$\langle \tilde{m}\rangle \sim \epsilon^{1/2}$ in regime II. However,
they begin to deviate from the straight line systematically for $N\ge
4096\times 10^3$ as the scaling variable increases. This is due to the
crossover to regime I.

Finally, we test the scaling form of Eq.~(\ref{fss-sf-2}), which is valid in
regimes I and II. Figure~\ref{fig5}(d) shows the scaling plot of
$\langle\tilde{m}\rangle N^{\beta/\bar{\nu}_*}$ against the scaling variable
$|\epsilon|N^{1/\bar{\nu}_*}$. As expected, we do not have a data collapse
for $\epsilon\le 0$. The data in regimes I and II do not collapse well
either. The order parameter scales as
$\langle\tilde{m}\rangle \sim \epsilon^{1/2}$ in regime I and then
$\langle\tilde{m}\rangle \sim \epsilon^{1/(\lambda-1)} =\epsilon^1$ in
regime II. Comparing the numerical data with the straight lines of slopes of
$1/2$ and $1$, one finds the signature of the crossover for
$N\ge 4096\times 10^3$. This suggests that the poor data collapse may be
due to a finite size effect. The system does not reach the scaling
regime I even at $N=32~768\times 10^3$ yet.

\begin{figure}[t]
\includegraphics[width=\columnwidth]{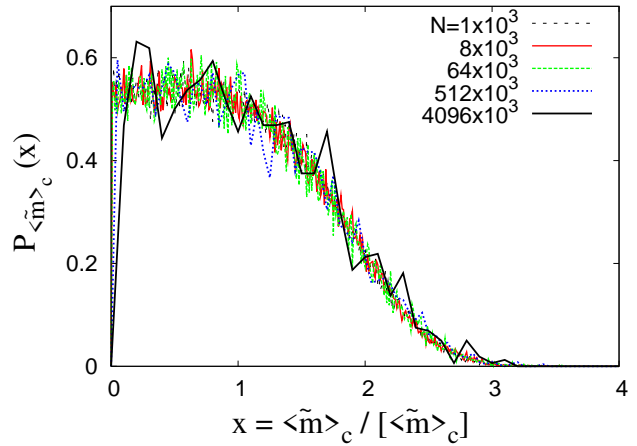}
\caption{(Color online) The histogram for $\langle\tilde{m}\rangle_c /
[\langle\tilde{m}\rangle_c]$, the order parameter at $\epsilon=0$
normalized with the ensemble average,
where network parameters are $\lambda=4.0$, $k_0=3$, $d=1$,
and $\omega=\omega_{\rm nat}=3$.
The curves from different values of $N$ collapse onto a single curve.}
\label{fig6}
\end{figure}

\section{Sample-to-Sample Fluctuations}
\label{sec5}

In the previous section, we tested the FSS theory for the power-law degree
distributions generated deterministically from Eq.~(\ref{det_way}).
The other way is to draw {\em probabilistically} $N$ values of the degree
independently in accordance with the target distribution function $P(k)$.
This is adopted in the configuration model~\cite{Newman2001,CBP-S2005}.
In the probabilistic method,
the degree sequence varies from sample to sample, hence an ensemble average
is necessary. One interesting issue is whether physical quantities have
the self-averaging property~\cite{Aharony1996} against the sample-to-sample
fluctuations.
For finite systems, a sample with $\{k_i\}=\{k_1,\ldots,k_N\}$ drawn
probabilistically may show the degree distribution
$\tilde{P}(k)=\sum_{i} \delta_{k,k_i}/N$, which deviates from the target
distribution function $P(k)$. Then, it follows that
the degree moments $z_n=\sum_i k_i^n/N$ show the sample-to-sample fluctuations,
purely from the sampling disorder.

Using the same techniques used in our previous publication for the CP model~\cite{JDNoh2009}
(see Sec. V therein), it is straightforward to show that
the relative fluctuation $R_n$ is given by
\begin{equation}
R_n\equiv\frac{[z_n^2]-[z_n]^2}{[z_n]^2}
=\frac{1}{N}\left(\frac{\langle 2n\rangle_0}{\langle n\rangle_0^2}-1\right),
\end{equation}
where $[\cdots]$ denotes the sample~(disorder) average and
$\langle n \rangle_0\equiv \sum_k k^n P(k)$. For exponential networks,
all $\langle n \rangle_0$ are finite, so all degree moments $z_n$ are strongly self-averaging
($R_n\sim N^{-1}$)~\cite{Aharony1996}.

In the SF networks with $P(k)$ given in Eq.~(\ref{p_k}),
\begin{equation}
\langle n\rangle_0 \sim \left\{
\begin{array}{lll}
  N^{(n-\lambda+1)/\omega} & \mbox{for}& n>\lambda-1 \\ [2mm]
  \log N & \mbox{for}& n=\lambda-1 \\ [2mm]
 \mathcal{O}(1) &\mbox{for}& n<\lambda -1,
\end{array} \right.
\label{n0}
\end{equation}
which leads to
\begin{equation}
R_n\sim \left\{
\begin{array}{lll}
 N^{-1+(\lambda-1)/\omega}  & \mbox{for}& n>\lambda -1 \\ [2mm]
 N^{-1+(2n-\lambda+1)/\omega} & \mbox{for}& (\lambda-1)/2 <n < \lambda-1 \\ [2mm]
 N^{-1} &\mbox{for}& n<(\lambda-1)/2,
\end{array} \right.
\label{Rn}
\end{equation}
where there are log corrections at $n=\lambda-1$ and $(\lambda-1)/2$.
By definition, $R_n$ is strongly self-averaging for $n<(\lambda-1)/2$
and is weakly self-averaging for $n>(\lambda-1)/2$ except that
$R_n$ is not self-averaging only when $\omega=\omega_{\rm nat}$ for
$n>\lambda-1$. For example, $z_4$ is not self-averaging for $\lambda<5$
with the natural upper cutoff.

The relevant quantities involving the degree moments are
$\tilde{K}_c=z_1/z_2$, $a=z_1^2/z_2$, and $b=(z_1/z_2)^4 z_4$.
It implies that the critical point location and $a$ are strongly
self-averaging for $\lambda>5$ and at least weakly self-averaging
for $\lambda>3$. However, $b$ is not self-averaging for $3<\lambda<5$ with
$\omega=\omega_{\rm nat}$,
which determines the amplitude of the order parameter in various
scaling regimes~[see Eqs.~(\ref{m_critical}) and~(\ref{m_r2})].
Therefore, we expect widely scattered data for the order parameter,
depending strongly on sampled degree sequences,
for $3<\lambda<5$ with $\omega=\omega_{\rm nat}$.

Numerical data are presented to verify the non-self-averaging property of
the order parameter at $\epsilon=0$, $\langle \tilde{m}\rangle_c \sim (bN)^{-1/4}$~[see Eq.~(\ref{m_critical})],
 for the annealed SF networks for $\lambda=4$ with
$\omega=\omega_{\rm nat}=3$.  It should not be self-averaging
because it involves the parameter $b$.
We have measured the order parameter $\langle \tilde{m}\rangle_c$
in many samples and constructed a histogram of the quantity
$\langle \tilde{m}\rangle_c / [ \langle \tilde{m}\rangle_c ]$,
the order parameter normalized with its mean values.

Figure~\ref{fig6} presents, thus, the obtained histogram.
The histogram does not sharpen at all,
but collapses onto a single curve as $N$ increases. This proves
the non-self-averaging property.

\section{Summary and discussion}
\label{sec6}

We have investigated the FSS of the Ising model
on annealed networks. The model is mapped to the Ising model on a globally
connected network with heterogeneous couplings, which allows us to derive the free-energy density
as a function of the magnetic order parameter $\tilde{m}$.
Using the free energy density function, the scaling functions for $\tilde{m}$
and the zero-field susceptibility $\tilde{\chi}$ are also derived.

For the networks with exponentially bounded degree distributions and
power-law degree distributions with $\lambda>5$, the FSS forms are given in
Eqs.~(\ref{m_fss_form}) and (\ref{sus_scaling_form}).
The critical exponents for the
magnetization and the susceptibility are given by $\beta=1/2$ and
$\gamma=1$, respectively. The FSS exponent is given by $\bar{\nu}=2$, with
which the scaling variable for the FSS is given by
$\epsilon N^{1/\bar{\nu}}$. The scaling behaviors in the critical
regime~[$\epsilon N^{1/\bar{\nu}} = \mathcal{O}(1)$],
in the supercritical regime~($\epsilon \gg N^{-1/\bar{\nu}}$),
and in the subcritical regime~($\epsilon \ll-N^{-1/\bar{\nu}}$)
are summarized in Figs.~\ref{fig1} and \ref{fig2}.

For power-law degree distributions with $3<\lambda<5$, the degree cutoff
$k_c \sim N^{1/\omega}$ matters and there exist two distinct scaling
variables $\epsilon N^{1/\bar{\nu}_c}$ with $\bar{\nu}_c =
2/[1-(5-\lambda)/\omega]$ and $\epsilon N^{1/\bar{\nu}_*}$ with
$\bar{\nu}_* = \omega/(\lambda-3)$ when $\omega>\omega_{\rm nat}=\lambda-1$. At
$\omega=\omega_{\rm nat}$, the two scaling variables merge into a single
one. The scaling behaviors in the supercritical
regime I~($\epsilon \gg N^{-1/\bar{\nu}_*}$), the intermediate regime
II~($N^{-1/\bar{\nu}_c} \ll \epsilon \ll N^{-1/\bar{\nu}_*}$),
the critical regime III~[$\epsilon N^{1/\bar{\nu}_c} = \mathcal{O}(1)$],
and the subcritical regime IV~($\epsilon\ll -N^{1/\bar{\nu}_c}$) are
summarized in Fig.~\ref{fig3}.
The crossover from regime I to II is originated from the finiteness of
the degree cutoff $k_c$, while the critical FSS in regime III is from
the finiteness of both $k_c$ and $N$.

The CP on the annealed SF network studied in Refs.~\cite{Boguna2009,JDNoh2009}
is also characterized with the two $\omega$-dependent FSS exponents when
$2<\lambda<3$ and $\omega>\omega_{\rm nat}$. The similarity between the
equilibrium Ising model and the nonequilibrium CP suggests that the
two-parameter scaling is a generic feature of critical phenomena in
annealed scale-free networks.

Extensive studies during the last decade have revealed that critical
phenomena on quenched networks and annealed networks are characterized with
the same set of bulk critical exponents such as the order parameter exponent
and susceptibility exponent. However, they display distinct FSS behaviors.
Annealed networks are characterized with two FSS exponents, which depend on
$\lambda$ and $\omega$. In comparison to annealed networks, quenched
networks have a quenched disorder in structure. Besides, dynamic degrees of
freedom on quenched networks have finite correlations. It is another big
challenge to understand how these two ingredients cause the distinct
FSS behaviors, some of which is under investigation~\cite{FSS-full}.

\acknowledgments
This work was supported by the BK21 project, Acceleration Research
(CNRC) of MOST/KOSEF, and Korea Research Council of Fundamental Science \&
Technology (S.H.L., M.H., and H.J.). Computation was carried out using KIAS
supercomputers.

\appendix
\section{scaling functions}\label{sec:app}

From the free energy density function in Eq.~(\ref{freg}) for the exponential networks
and also the SF networks with $\lambda>5$, one can easily derive $\langle\tilde{m}\rangle$ for
small $\tilde{h}$ as
\begin{equation}
\langle\tilde{m}\rangle\simeq \left(\frac{12}{bN}\right)^{1/4}
\frac{U(1/2,r)+a'N^{3/4} \tilde{h} U(3/4,r)}{U(1/4,r)+a'N^{3/4} \tilde{h} U(1/2,r)},
\end{equation}
where $a'=(12/b)^{1/4} a$, $r=(3a^2/b)^{1/2}\epsilon N^{1/2}$, and
$$U(s,r)=\int_0^\infty dy\ y^{s-1} \exp (-y + r \sqrt{y}).$$

With this, we find
the order parameter scaling at $\tilde{h}=0$ as
\begin{equation}\label{m_fss_form}
\langle\tilde{m}(\epsilon,N)\rangle=N^{-\beta/\bar\nu} \tilde{\psi}
(\epsilon N^{1/\bar\nu};a,b),
\end{equation}
where $\beta=1/2$, $\bar\nu=2$, and
\begin{equation}\label{m_s_f}
\tilde{\psi}(x;a,b)= \left(\frac{12}{b}\right)^{1/4}
\frac{U(1/2,r_0 x)}{U(1/4, r_0x)},
\end{equation}
with $r_0 = (3a^2/b)^{1/2}$ and $x=\epsilon N^{1/\bar{\nu}}$.

The zero-field susceptibility
$\tilde{\chi}=(\partial \langle\tilde{m}\rangle)/(\partial \tilde{h})|_{\tilde{h}=0}$
is
\begin{equation}\label{sus_scaling_form}
\tilde{\chi}(\epsilon,N)=N^{\gamma/\bar\nu} \tilde{\phi}
(\epsilon  N^{1/\bar\nu};a,b),
\end{equation}
where $\gamma=1$ and
\begin{equation}\label{sus_s_f}
\tilde{\phi} (x;a,b)=a\left(\frac{12}{b}\right)^{1/2}\left[ \frac{U(3/4,r_0 x)}{U(1/4, r_0x)}
-\frac{U^2(1/2,r_0 x)}{U^2(1/4, r_0x)}\right].
\end{equation}

Using the properties of the  function $U(s,r)$ such as
\begin{equation}
U(s,r) \simeq \left\{
\begin{array}{ll}
2 \Gamma (2s)\ (-r)^{-2s} &  (r\rightarrow -\infty) \\ [2mm]
\Gamma (s) &  (r=0) \\ [2mm]
2\pi^{\frac{1}{2}}e^{\frac{r^2}{4}}  \left(\frac{r}{2}\right)^{2s-1} \left[ 1+\frac{4(s-1)(s-2)}{r^2} \right]
 & (r\rightarrow \infty),
\end{array} \right.
\end{equation}
one can show
$\tilde\psi (x)\simeq \sqrt{3a/b}\ x^{1/2}$ or
$\sqrt{2/(\pi a)} (-x)^{-1/2}$  for $x\rightarrow \pm\infty$,
and $\tilde\phi (x)\simeq (2x)^{-1}$ or
$(1-2/\pi) (-x)^{-1}$ for $x\rightarrow \pm\infty$.
We remark that the usual magnetization
and the magnetic susceptibility become $m\simeq a\langle\tilde{m}\rangle$ and
${\chi}\simeq a\tilde{\chi}$.

For the SF networks with $3<\lambda<5$, the scaling function $\tilde{\psi}_c (x)$
for the order parameter near $\epsilon\approx 0$ in Eq.~(\ref{fss-sf-1}) behaves
in the same way as the above $\tilde\psi (x)$ except for replacing
$\beta$ by $\tilde\beta$, $\bar\nu$ by $\bar{\nu}_c$, and $b$ by $b_0$. For example,
$r$ becomes $r=(3a^2/b_0)^{1/2}\epsilon N^{1/\bar\nu_c}$. The susceptibility
scaling function also changes in  the same way.


\begin{thebibliography}{99}
\bibitem{exp1} The maximum value of $p_{ij}\sim N^{2/\omega-1}$ with
$k_c\sim N^{1/\omega}$ and $\omega\ge\lambda-1$ in the SF networks with  $P(k)\sim k^{-\lambda}$.
Therefore, the condition that $p_{ij} \ll 1$ for large $N$ is valid for general $\omega$ only when $\lambda>3$.
\end{thebibliography}
\end{document}